\documentclass[amsmath,amssymb,superscriptaddress,onecolumn]{revtex4-2}

\usepackage{graphicx}
\usepackage{dcolumn}

\def\be{\begin{equation}}
\def\ee{\end{equation}}
\def\beq{\begin{eqnarray}}
\def\eeq{\end{eqnarray}}

\usepackage{bm}
\usepackage{graphicx}
\usepackage{dcolumn}
\usepackage{amsmath}
\usepackage[latin1]{inputenc}
\usepackage{graphicx, psfrag}
\usepackage{amssymb}
\usepackage[colorlinks=true, citecolor=blue, urlcolor = blue, linkcolor= red, bookmarks=true]{hyperref}
\usepackage{float}
\usepackage{amsmath}
\usepackage{amsfonts}
\usepackage{dcolumn}
\usepackage{hyperref}
\usepackage{subfigure}
\usepackage{pgfplots}
\usepackage{epstopdf}
\usepackage{xcolor}
\usepackage{booktabs}
\usepackage{csquotes}

\begin{document}


\title{Is dark energy necessary for the sustainability of traversable wormholes?
}


\author{Ayan Banerjee} \email[]{ayanbanerjeemath@gmail.com}
\affiliation{Astrophysics Research Centre, School of Mathematics, Statistics and Computer Science, University of KwaZulu--Natal, Private Bag X54001, Durban 4000, South Africa}

\author{Sudan Hansraj} 
\email[]{hansrajs@ukzn.ac.za}
\affiliation{Atrophysics Research Centre, School of Mathematics, Statistics and Computer Science, University of KwaZulu--Natal, Private Bag X54001, Durban 4000, South Africa}

\author{Anirudh Pradhan}
\email{pradhan.anirudh@gmail.com}
\affiliation{Centre for Cosmology, Astrophysics and Space Science, GLA University, Mathura-281 406, Uttar Pradesh, India}

\author{Abdelghani Errehymy} 
\email[]{abdelghani.errehymy@gmail.com}
\affiliation{Atrophysics Research Centre, School of Mathematics, Statistics and Computer Science, University of KwaZulu--Natal, Private Bag X54001, Durban 4000, South Africa}


\date{\today}

\begin{abstract}
In the standard approach to studying wormhole geometry, the presence of dark energy is unavoidable to ensure traversability. The dark energy provides the negative gravity effect to keep the throat open. The question we analyse is whether the same can be achieved without dark energy. It turns out that if we couple the trace of energy-momentum with the standard Einstein-Hilbert Lagrangian and utilise a sppecific equation of state then dark energy may be obviated.  The Casimir stress energy is known to result in the violation of the null energy condition (NEC) on the energy momentum tensor. This phenomenon makes such an equation of state (EoS) an ideal candidate to generate traversable wormhole (WH) geometries. The laboratory proven phenomenon provides a mechanism to sustain an open WH throat without having to appeal to dark energy.  We generate two classes of WH solutions with this  in $f(R,T)$ gravity theory, where $R$ represents the Ricci scalar and $T$ is the trace of the stress-energy tensor.
For the background geometry we choose a static and spherically symmetric metric, and derive the field equations for exact WH solutions. For the specific choice of the Casimir EoS relating the energy-momentum tensor components [ Kar and Sahdev: Phys. Rev D {\bf 52} 2030 (1995)] and different choices for redshift functions, we determine the WH geometry completely. The obtained WH solutions violate the  NECs and all qualitative constraints demanded of physically realisable WHs are satisfied. This is demonstrated via graphical plots for a suitably chosen range of values of the $f(R,T)$ coupling parameter. Furthermore, our study involved an investigation into the repulsive effect of gravity, which revealed that its presence leads to a negative deflection angle for photons traveling along null geodesics. Notably, we observed a consistent pattern of negative values for the deflection angle across all values of $r_0$ in the three scenarios considered, thus indicating the clear manifestation of the repulsive gravity effect. All of this is possible without invoking the existence of dark energy. 

\end{abstract}


\maketitle


\section{Introduction}\label{intro:Sec}
 
The concept of a WH as a tunnel-like structure in the spacetime continuum connecting two regions of a manifold or two separate universes was initiated by Einstein and Rosen \cite{Einstein:1953tkd} and developed by Morris and Thorne (MT) \cite{morris}. The MT proposal was beset with some problems in the form of the throat not remaining open long enough to allow even a photon to pass through. However, it was conjectured by \cite{morris} that suitable geometries could be found such that even humans could pass through WHs. These are called traversable WHs. Theoretically it is even possible to entertain the idea of time travel.  Bronnikov \cite{bronnikov} considered the novel construction of cylindrically symmetric WHs where one end existed in a 4 dimensional spacetime and the other opened up into a 6 dimensional geometry. He also discussed possible observable features of such WHs to aid the search for the same.   It is also required that  WHs should be horizon-free and singularity free everywhere \cite{dotti}.

A central requirement for a WH throat to form in general relativity is the violation of the NEC and consequently the existence of exotic types of matter fields is implied. Physically this means that a negative density facilitates a repulsive gravitational force which is required to keep the WH throat open.  At first MT WHs were studied and the idea was to find solutions and to check that the NEC was violated later. Then Visser \cite{visser5,visser6} contemplated the idea of commencing by gluing together  two spacetime manifolds that are asymptotically flat and using the junction conditions to establish conditions for the existence of WHs. These conditions invariably amounted to a violation of the NEC at the outset.  For example, see the work of Kar and Sahdev \cite{Kar:1995vm} who solved the Einstein equations after prescribing an  with trace-free energy momentum. This   corresponds to the  Casimir stress energy and has the feature of necessary NEC violation. This makes this  ideal in the study of WHs. (Note that fluids with the Casimir  will be considered as exotic in this work. Generally exotic fluid is taken to mean dark energy but that is not the case in this article.) Evidently this stipulation has not been   intensively studied in the literature.  Kar and Sahdev \cite{Kar:1995vm} reduced the defining equation to a nonlinear second order differential equation  after a restrictive prescription of the redshift function {\it{de facto}}  the temporal potential. The choice is made such that a negative gradient was evident and no singularities or horizons should emerge. Following the success of this process the same authors proceeded to join two spacetime manifolds as per the Visser scheme and generated two equations from the junction conditions which they solved and constructed a viable WH solution. This idea of using the Casimir stress energy has ostensibly not been applied in modified gravity theories. This is surprising since NEC violation is guaranteed and this is highly desirable in modelling WHs. This strongly motivates our consideration of this  in constructing viable WHs in $f(R, T)$ theory. 

 The Casimir effect is well known in standard physics as an attractive force between neutral parallel plates in a vacuum. This force is a result of the vacuum energy of quantum fields. It is a pure quantum effect. The Casimir effect is one of the few phenomena in physics that can create an observable effect of negative energy density. Between the plates, the energy density can become effectively negative compared to the free vacuum. This unusual behaviour makes it a good candidate for an actual mechanism to generate a negative energy. The Casimir effect is significant in WH physics primarily as a theoretical example of how negative energy densities, necessary for the existence of a stable WH, might be achieved in a controlled physical setting. Garattini \cite{garattini} analysed the Casimir effect for WH physics and developed a viable traversable WH model assuming such an .

While the focus of this investigation revolves around the effect of the Casimir stress energy which has the characteristics of exotic matter but which is a laboratory proven extraordinary feature of standard physics, the question of whether exotic matter fields are compulsory in order to ensure an open WH throat is certainly worth considering. In the present context one could ask whether the modifications to the field equations introduced by the $f(R, T)$ Lagrangian, specifically the addition of a scale dependent trace of the energy momentum tensor, could alter the geometry sufficiently thus obviating the need for exotic matter. Sporadic research along these lines have appeared in the literature in the case of $f(R, T)$ gravity. For instance in \cite{zubair0} some models of WHs not demanding exotic matter are reported.  We point out that the possibility of avoiding exotic matter to keep WH throats open has been addressed in several other theories. For example, through the involvement of a scalar field in Brans-Dicke gravity, exotic fields are not required to sustain the geometry of open WHs \cite{garcia,papan,Anchordoqui:1996jh}.  Bahamonde {\it{et al}} \cite{bahamonde} constructed $f(R)$ gravity models for the universe with WHs that do not require exotic matter in general however  the general relativity case $f(R) = R$ cannot avoid exotic matter \cite{bahamonde}.  Zubair {\it{et al}} \cite{zubair} commenced a study of WHs in $f(R, T)$ gravity but then specialised to a quadratic form of $f(R)$ theory and concluded from their models that exotic matter was not required to sustain the WH. Despite these results it remains an interesting question to consider that is whether mechanisms in standard physics exist that can sustain WHs. A recent investigation by Nojiri, Odintsov, et al. \cite{Nojiri:2024dde, Elizalde:2023rds} has introduced a novel approach to the study of WHs in modified gravity. A comprehensive study presented in \cite{Richarte:2007zz, Richarte:2010bd, Cuyubamba:2018jdl, Boehmer:2012uyw, Bolokhov:2012kn, Capozziello:2012hr, Richarte:2013lua, Bahamonde:2016jqq, DiGrezia:2017daq, Amir:2018pcu, Javed:2019qyg, Antoniou:2019awm, DeFalco:2020afv, Capozziello:2020zbx, Benavides-Gallego:2021lqn, Karakasis:2021tqx, DeFalco:2021ksd, DeFalco:2021klh, Abdulxamidov:2022ofi, Capozziello:2022zoz, DeFalco:2021btn, Chanda:2021dvc,Errehymy:2024yey,Errehymy:2023rsm,Errehymy:2023rnd,Errehymy:2024cgy,Mustafa:2024jsv,Mustafa:2023ojf,Ditta:2021uoe} provides detailed information on various examples of traversable WH solutions in different modified gravity theories, highlighting the exciting works of literature in this field. For this reason we study this possibility. 

But why should $f(R, T)$ theory be of interest amongst the many alternatives to general relativity? It is well known that $f(R, T)$ theory has succeeded in explaining a cornerstone problem in gravitation namely the observed accelerated expansion of the universe which the standard theory cannot deal with without having to appeal to dark matter and dark energy \cite{Harko:2011kv}. Moreover, $f(R, T)$ theory admits astrophysical models that comply with the elementary requirements for stellar distributions \cite{hans-ban-prd}. One major trade-off though with $f(R, T)$ is that energy conservation is sacrificed. This is also true for other viable theories such as unimodular gravity \cite{ellis1,ellis2,hans-ellis}, Rastall theory \cite{rastall1,rastall2,hans-ban-chan,hans-ban-mpla} and Weyl quadratic theory \cite{yang,weyl2,scholz} and its generalisation by Dirac \cite{dirac1,dirac2,rosen,israelit}.  In the context of WHs the loss of energy conservation is not necessarily a negative feature given that NEC violation is demanded and loss of energy into a WH is probable.

The question of whether WHs are realistic physical entities is interesting. Specifically we are interested in what observational signatures would indicate the presence of WHs. Some observational  protocols have recently emerged.  For instance Piotrovich {\it{et al}} \cite{piotrovich} expressed the view that active galactic nuclei  are WH mouths instead of supermassive black holes as is the prevailing view. They demonstrated  that accreting flows generate collisions inside WHs and  gamma ray radiation is the consequence.  In their view  if such radiation were detected it would serve as solid proof for the existence of WHs.   It is therefore not surprising that WHs have attracted considerable attention in the literature. Some configurations that have been studied include  a Chaplygin gas  and   dark matter as the source of phantom energy  \cite{sushkov,lobo5,lobo6,peter,jamil1,jamil2,lobo7,cataldo,parsaei,Kuhfittig:2018hts}. Geometric effects such as  torsion, nonmetricity or higher curvature and higher dimensional  terms may also account for exotic matter in some modified theories of gravity. Additionally Lobo  \cite{lobo} and Visser \cite{visser} have also considered more factors that could explain the exotic matter.

   Deng and Meng \cite{wang} constructed traversable WHs by assuming the presence of dark energy. In this work, they appealed to certain astrophysical observations to dismiss  other WH solutions. They produced  6 models one of which was asymptotically flat while the other 5 was built from matter  spatially distributed near the throat of the WH. Some researchers have found novel ways to obviate exotic matter \cite{kanti1, kanti2}. Blazquez-Salcedo {\it{et al}} \cite{blaz} generated a class of singularity-free traversable WHs in the Einstein-Dirac-Maxwell theory without resorting to exotic matter. They make use  of two massive fermions that admit asymptotically flat WH solutions. A physically reasonable traversable WH model was developed by \cite{konoplya} without the mirror symmetry of \cite{blaz}.  Another viable  approach is to use the  thin-shell formalism \cite{israel,darmois}. In this method Lobo  \cite{lobo9} explains that suitable patches of a manifold could be pasted together such that the violation of the NEC occurs only on a thin shell.

WHs exhibit different properties depending on the gravitational field theory being used. In some theories exotic matter is not required for WHs to exist. This has been shown by   Pavlovic and Sossich \cite{pavlovic} and Harko et al. \cite{harko}  in the case of   $f(R)$ theory.    De Benedictis and Horvat \cite{benedictis} demonstrated the throat of a WH in the context  of  $f(R)$ gravity. Sharif and Zahra \cite{sharif} investigated the role of pressure anisotropy in the formation of WHs in the presence of an  while Eiroa and Aguirre \cite{eiroa} studied thin-shell Lorentzian WHs in  $f(R)$ theory.   Traversable WHs  in $f(R)$ gravity were considered by  Mazharimousavi and Halilsoy \cite{mazhar}.  Godani and Samanta \cite{godani}  examined the role of various redshift functions on on WH structure and generated some new solutions.\\

\smallskip

\smallskip

The outline of this article is as follows. In Sec. \ref{sec2}, we begin by briefly reviewing the field equations of $f(R,T)$ theory. In Sec. \ref{sec3}, we have written the corresponding field equations for spherically symmetric and static spacetimes in $f(R,T)$ gravity. In the same section we present an EoS that reduces to a traceless energy momentum tensor. The structure equation for standard energy conditions is presented in Sec. \ref{sec4}.  We will then construct analytic WH solutions for different kinds of redshift and shape functions in Sec. \ref{sec5}. In Sec. \ref{sec6} we have investigated the
deflection angle of a light ray passing close to the wormhole. We end with a conclusion in Sec. \ref{sec7}.

\section{Action and equations of the $f(R,T)$ theory}
\label{sec2}
Various modified gravity theories have been proposed to extend generaal relativity.  The $f(R,T)$ gravity model \cite{Harko:2011kv}
which comes from the combination of curvature-mater coupling theory has many applications in describing the current Universe. Let us start from the $f(R,T)$ action given by
\begin{equation}\label{1}
    \mathcal{A} = \frac{1}{16\pi}\int f(R,T)\sqrt{-g}d^4x + \int\mathcal{L}_m \sqrt{-g}d^4x ,
\end{equation}
where $g$ is the determinant of the metric $g_{\mu\nu}$, and $\mathcal{L}_m$ denotes the action of matter. The $f(R,T)$ 
is an arbitrary function of the Ricci scalar $R$ (Ricci scalar) and $T$ (trace of the energy-momentum tensor). 

Variations of the action (\ref{1}) with respect to the metric $g_{\mu\nu}$ leads to the modified field equations of the $f(R,T)$ gravity, yielding 
\begin{align}\label{2}
    f_R(R,T) R_{\mu\nu} - \dfrac{1}{2}f(R,T) g_{\mu\nu} + [g_{\mu\nu}\square - \nabla_\mu\nabla_\nu] f_R(R,T)  = 8\pi T_{\mu\nu} -(T_{\mu\nu} + \Theta_{\mu\nu})f_T(R,T) ,
\end{align}
where we have defined $f_R \equiv \partial f/\partial R$, $f_T \equiv \partial f/\partial T$, $\square \equiv \nabla_\mu\nabla^\mu$ is the d'Alembertian operator with covariant derivative represented by $\nabla_\mu$. Here, the auxiliary tensor $\Theta_{\mu\nu}$ defined in terms of the variation of $T_{\mu\nu}$ as
\begin{align}\label{eq3}
    \Theta_{\mu\nu} \equiv g^{\alpha\beta}\frac{\delta T_{\alpha\beta}}{\delta g^{\mu\nu}}  =  -2T_{\mu\nu} + g_{\mu\nu}\mathcal{L}_m - 2g^{\alpha\beta} \frac{\partial^2\mathcal{L}_m}{\partial g^{\mu\nu} \partial g^{\alpha\beta}} .
\end{align}

Taking the trace of the field equations (\ref{2}),  one can get a relation between $R$ and $T$ of the form:
\begin{align}\label{4}
    3\square f_R(R,T) &+ Rf_R(R,T) - 2f(R,T) = 8\pi T - (T+\Theta)f_T(R,T) ,
\end{align}
where $\Theta= g^{\mu\nu}\Theta_{\mu\nu}$ is the trace of $\Theta_{\mu\nu}$. Finally, we obtain the covariant derivative of the stress-energy tensor and using the field equations (\ref{2}), which yield \cite{BarrientosO:2014mys}
\begin{align}\label{eq5}
    \nabla^\mu T_{\mu\nu} =\ \frac{f_T(R,T)}{8\pi - f_T(R,T)}\bigg[ (T_{\mu\nu} + \Theta_{\mu\nu})\nabla^\mu \ln f_T(R,T)  + \nabla^\mu\Theta_{\mu\nu} - \frac{1}{2}g_{\mu\nu}\nabla^\mu T \bigg] .
\end{align}

Interestingly, the Eq. (\ref{eq5}) is not conserved and this is due to the interaction between curvature and matter sectors. Moreover, non-conservation leads to nongeodesic motion of particles. In the case of $f(R,T)$ gravity the functional form of $f$ is not unique. Many $f(R,T)$ models exhibit interesting features, but here we continue our investigation by considering the simplest functional 
form of $f(R,T)= R+ 2\beta T$ \cite{Harko:2011kv} i.e., the usual Einstein-Hilbert term plus a $T$ dependent function $f(T )$. This model has been widely applied in astrophysical scenarios and also
explains the accelerated expansion of the Universe in an elegant manner. Using the particular form function, we rewrite the Eqs.~(\ref{2}), (\ref{4}) and (\ref{eq5}) as
\begin{align}
    G_{\mu\nu} &= 8\pi T_{\mu\nu} + \beta Tg_{\mu\nu} - 2\beta(T_{\mu\nu} + \Theta_{\mu\nu}) ,   \label{eq6}   \\ 
    R &= -8\pi T - 2\beta(T- \Theta) ,   \label{7}    \\
    \nabla^\mu T_{\mu\nu} &= \frac{2\beta}{8\pi - 2\beta} \left[ \nabla^\mu \Theta_{\mu\nu} - \frac{1}{2}g_{\mu\nu}\nabla^\mu T \right] ,   \label{8}
\end{align}
where $G_{\mu\nu}$ is the Einstein tensor.

\section{The WH geometry and the field equations} \label{sec3}

To study the WH structure, we start by considering the static spherically symmetric
metric ansatz in Schwarzschild coordinates in the form 
\begin{equation}\label{metric}
ds^2= -e^{2\Phi(r)}dt^2+\frac{dr^2}{1-\frac{b(r)}{r}}+r^2(d\theta^2+\sin^2\theta d\phi^2),
\end{equation}
and this is in the form of the MT metric \cite{morris}. Here $\Phi(r)$ is called the redshift function and $b(r)$ is the shape function. The coordinate $r$ 
lies in  the range of $[r_0, +\infty)$ and $(-\infty, r_0]$ with a minimum value at $b(r_0)= r_0$.  This can be interpreted as a `throat' of the WH with circumference of a circle  given by $2\pi r$. More precisely, the shape function should satisfy the flaring-out condition i.e.,  
$\frac{b(r)-rb^{\prime}(r)}{b^2(r)}>0$ \cite{morris} and this is provided by the mathematics of embedding. This implies another restriction $b^{\prime}(r_0) < 1$ at the throat. In addition,  the shape function also obeys $b(r)-r \leq 0$. Nevertheless $\Phi(r)$ must be finite everywhere to ensure the absence of an event horizon. 

Here, we adopt an anisotropic fluid form that describes the matter sector. Such a tensor can be written as 
\begin{equation}\label{eq7}
T_{\mu\nu}=(\rho+p_t)u_\mu u_\nu+ p_t g_{\mu\nu}-\sigma \chi_{\mu}\chi_{\nu},
\end{equation}
where $\rho = \rho(r)$ is the energy density, $p_r = p_r(r)$ and $p_t = p_t (r)$ are the radial and transverse pressures, respectively. The $u^\mu$ is the 4-velocity and $g_{\mu\nu}$ is the metric tensor. The anisotropy factor  is given by $\sigma = p_t-p_r$.

It is well known from the definition of the matter Lagrangian for isotropic/anisotropic fluid that the energy-momentum tensor 
is not unique. We could choose either $\mathcal{L}_m = \mathcal{P}$ or $\mathcal{L}_m = -\rho$.  In the present study, we consider $\mathcal{L}_m = \mathcal{P}$,  where $\mathcal{P} \equiv (p_r+ 2p_t)/3$ \cite{Deb:2018sgt,Maurya:2019sfm,Biswas:2020gzd}, which allows us to rewrite Eq. (\ref{eq3}) as $\Theta_{\mu\nu} = -2T_{\mu\nu} + \mathcal{P}g_{\mu\nu}$. By taking the covariant divergence of Eq. (\ref{8}) we obtain the result
\begin{align}
    \nabla^\mu T_{\mu\nu} &= \frac{2\beta}{8\pi + 2\beta} \left[ \nabla^\mu \left(\mathcal{P}g_{\mu\nu} \right) - \frac{1}{2} \nabla_\nu T \right] .   \label{13}
\end{align}
We note that the standard conservation equation does not hold for this theory. If we take the $\beta =0$, one can recover the conservation of energy-momentum tensor for the general relativistic
case. Finally, combining Eqs. (\ref{metric}), (\ref{eq7}) and the gravitational equation (\ref{eq6}), we reach the following expressions
  \begin{eqnarray}
      8\pi\rho + \beta\left[ 3\rho - p_r - \frac{2}{3}\sigma \right]&=& \frac{b^{\prime}}{r^{2}}   ,  \label{12}  \\
        8\pi p_r + \beta\left[ -\rho + 3p_r+ \frac{2}{3}\sigma \right]&=& 2\left(1-\frac{b}{r}\right)\frac{\Phi^{\prime}}{r}-\frac{b}{r^{3}} ,  \label{13}  \\
        8\pi p_t  + \beta\left[ -\rho + 3p_r+ \frac{8}{3}\sigma \right] &=& \left(1-\frac{b}{r}\right)\left[\Phi^{\prime\prime}+\Phi^{\prime 2}-\frac{b^{\prime}r-b}{2r(r-b)}\Phi^{\prime}-\frac{b^{\prime}r-b}{2r^2(r-b)} +\frac{\Phi^{\prime}}{r}\right]   ,  \label{14}
  \end{eqnarray}
where the prime stands for differentiation with respect to $r$. Finally, we have three independent equations (\ref{12})-(\ref{14}) 
with five unknown quantities, i.e., $\Phi(r)$, $b(r)$, $\rho(r)$, $p_r(r)$ and $p_t(r)$, respectively. Obviously this system is underdetermined and we are free to make any two assumptions to generate a physically reasonable WH solution. 

For the sake of simplicity, one can consider an  EoS of the form \cite{Anchordoqui:1996jh} 
\begin{eqnarray}\label{EoS1}
\rho(r)= \omega \left[p_r+2p_t\right].
\end{eqnarray}
With these considerations, one may write the EM tensor using Eq. (\ref{EoS1}) as $T= -\rho+p_r+ 2p_t=\rho(1-\omega)/ \omega$. In particular, the EoS  (\ref{EoS1})  reduces to a traceless EoS i.e., $T = 0$ for $\omega=1$, and the situation is related to the Casimir effect. By choosing this particular EoS, authors in \cite{Mehdizadeh:2015jra,Mehdizadeh:2016nna} have reported WH solutions in higher dimensional gravity theory. Most importantly this EoS has the feature of necessarily violating the NEC which is a necessity to maintain the throat of a WH in the open position due to the negative density inducing a repulsive gravitational effect.  In light of these considerations i.e. $\omega=1$, we endeavour to construct WH models in the context of $f(R,T)$ gravity.

\section{Energy Conditions} \label{sec4}
This section is devoted to discussing some properties of the stress-energy tensor that will be useful when analysing the energy conditions. As we  know the violation of NEC at least in a neighborhood is the most salient feature of these space-times from the perspective of GR \cite{morris}. As a result, we are searching for the existence of 
WH solutions in $f(R,T)$ theory with suitable energy conditions  such that  the matter field violates the NEC.  
Given the stress-energy tensor (\ref{eq7}), the NEC  is given by
\begin{equation}
\rho+p_i \ge 0,~\text{where}~ i = r, t,
\end{equation}
while the WEC asserts that
\begin{equation}
\rho \ge 0 ~\text{and} ~\rho+p_i\ge 0,
\end{equation}
and the expression for SEC implies
\begin{equation}
\rho +\sum_{i} p_i \ge 0 ~\text{and for each} ~\rho+p_i\ge 0.
\end{equation}
The above expressions will prove useful when searching for WH solutions with NEC violation preferably not from exotic matter such as dark energy.

\section{Solution generating technique} \label{sec5}

Here, we start this section by finding WH solutions with two  different kinds of redshift functions and one shape function taking into account that $f(R,T)= R+ 2\beta T$. For the WH to be traversable, we consider $\Phi^{\prime}(r)=0$ and $\Phi(r)= \log \left(1+\frac{r_0}{r}\right)$.  This choice guarantees that the redshift function is finite everywhere, and consequently the geometry is horizon free. Finally, we consider the  following shape function  $b(r) = \frac{r_0^2}{r}$, which ensures an asymptotically flat spacetime  satisfying the flaring-out condition at the WH throat. These assumptions simplify the field equations significantly and provide particularly intriguing solutions which we provide below.

\subsection{Solution for $\Phi^{\prime}(r)=0$} 

The simplest approach is to consider $\Phi^{\prime}(r) = 0$ implying WHs with zero tidal forces. In this context  Eqs. (\ref{12})-(\ref{14}) with the EoS (\ref{EoS1}) evaluate to the system  
\begin{eqnarray}
\frac{b'(r)}{r^2} &=&\frac{8}{3} \left[\beta +3 \pi ) (p_r+2 p_t\right],\label{eq19}\\
 -\frac{b(r)}{r^3} &=&\frac{4}{3} \left[(\beta +6 \pi ) p_r-\beta  p_t\right], \label{eq20} \\
 \frac{b-r b'}{2 r^3}&=&\frac{2}{3} \left[(\beta +12 \pi ) p_t-\beta  p_r\right]. \label{eq21}
\end{eqnarray}
From the system of  Eqs. (\ref{eq19})-(\ref{eq21}), we may express $p_r$ and $p_t$ in terms  of $b$ and $b^{\prime}$ in the form 
\begin{eqnarray}
 p_r(r)&=& \frac{\beta  r b'-4 (\beta +3 \pi ) b}{8 (\beta +3 \pi ) (\beta +4 \pi ) r^3}, \label{eq22}\\
 p_t(r)&=& \frac{(\beta +6 \pi ) r b'+2 (\beta +3 \pi ) b}{8 (\beta +3 \pi ) (\beta +4 \pi ) r^3}. \label{eq23} 
\end{eqnarray}
Substituting $p_r$ and $p_t$ given in (\ref{eq22}) and (\ref{eq23})
in Eq. (\ref{eq21}), we obtain the shape function $b(r)=C$. Using the condition $b(r_0 )= r_0$ one can write $b(r)= r_0$, which is constant. Substituting those values into the Eqs. (\ref{eq19})-(\ref{eq20}), we get the components of the energy-momentum tensor 
 $\rho(r)=0$,  $p_r(r) = -\frac{r_0}{2 (\beta +4 \pi ) r^3}$,  $p_t(r) = \frac{r_0}{4 (\beta +4 \pi ) r^3}$.  This is a physically untenable situation since the energy density is zero while $p_r = -2p_t$. Thus, considering $\Phi^{\prime}(r)=0$ is not suitable for describing  spherically symmetric WH solutions in $f(R,T)$ gravity. Interestingly,  zero-tidal-force WHs with isotropic pressure and a single matter source was not  possible in GR also, see Ref. \cite{Kar:1995vm}. 

\subsection{Solution for $\Phi(r) = \log \left(1+\frac{r_0}{r}\right)$} 

The second case which we are going to consider is logarithmic: $\Phi(r) = \log \left(1+\frac{r_0}{r}\right)$.  In
this  model, the field equations (\ref{12})-(\ref{14}) with the EoS (\ref{EoS1}), reduce to the system 
\begin{eqnarray}
 \frac{b'}{r^2} &=& \frac{8}{3} \left[\beta +3 \pi \right] \left[p_r+2 p_t\right],\label{eq25}\\
 \frac{b (r_0-r)-2 r r_0}{r^3 (r+r_0)} &=& \frac{4}{3} \left[(\beta +6 \pi ) p_r-\beta  p_t\right], \label{eq26} \\
\frac{r \left[2 r_0-r b'\right]+b (r-2 r_0)}{2 r^3 (r+r_0)} 
 &=&\frac{2}{3} \left[(\beta +12 \pi ) p_t-\beta  p_r\right]. \label{eq27} \nonumber\\ 
\end{eqnarray}

Now, the first two equations gives the expression for $p_r$ and $p_t$ in terms  of $b$ and $b^{\prime}$, 
\begin{align}
 p_r(r)&=\frac{\beta  r (r+r_0) b'-4 (\beta +3 \pi ) (b (r-r_0)+2 r r_0)}{8 (\beta +3 \pi ) (\beta +4 \pi ) r^3 (r+r_0)},\label{eq27}\\
 p_t(r)&= \frac{(\beta +6 \pi ) r (r+r_0) b'+2 (\beta +3 \pi ) (b (r-r_0)+2 r r_0)}{8 (\beta +3 \pi ) (\beta +4 \pi ) r^3 (r+r_0)},\label{eq28}
\end{align}

\begin{figure*}[h]
    \centering
    \includegraphics[width = 5.8 cm,height=5.8cm]{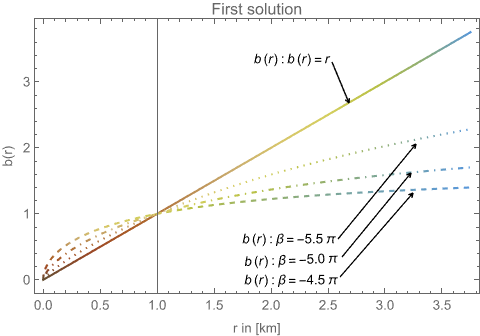} 
    \includegraphics[width = 5.8 cm,height=5.8cm]{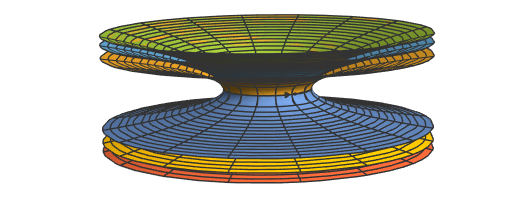} 
    \includegraphics[width = 5.8 cm,height=5.8cm]{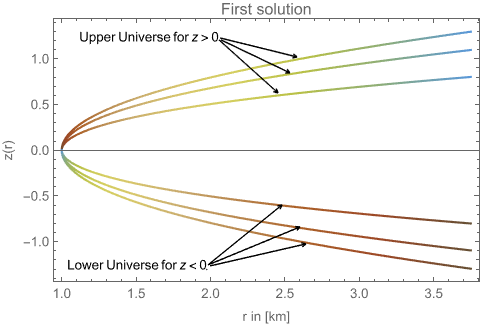} 
    \caption{Left panel: The behavior of $b(r) = r$ (solid black), and other curves are plotted for $b(r)$ given in Eq. (\ref{bb3}) which shows $b(r) < r$. Here,  we consider a specific redshift function $\Phi(r) = \log \left(1+\frac{r_0}{r}\right)$ and plot $b(r)$ for $\beta =-4.5 \pi$ (solid), $\beta =- 5\pi$(dashed) and $\beta =-5.5 \pi$ (dot-dashed), respectively. Middle panel:  Using the Eq. (\ref{bb3}) and Eq. (\ref{embed}), we plot the embedded diagram, wich provide valuable insights into the geometry and topology of the WH configurations. Right panel: Using the Eq. (\ref{bb3}) and Eq. (\ref{embed}), we plot the embedded surface $z(r)$, which provide valuable insights into the geometric structure and characteristics of the embedded surface. In all cases, we consider $\omega=1$, which reduces to a traceless EMT, with $T = 0$.  See the text for specific choices of the considered parameters. }  \label{fig1}
\end{figure*}

Plugging the expressions for $p_r$ and $p_t$ in Eq. (\ref{eq27}), we arrive at the following master equation
\begin{eqnarray} \label{eq29}
 b'(r)= -\frac{(\beta +3 \pi ) r_0 b}{r (\beta  r+3 \pi  (2 r+r_0))}.
\end{eqnarray}
Equation (\ref{eq29}) is routinely solved in the form 
\begin{eqnarray} \label{bb3}
b(r)= r_0 \left(\frac{  6\pi r+3 \pi r_0+ \beta r}{(\beta +9\pi)r }\right)^{\frac{3\pi+\beta}{3\pi}},
\end{eqnarray}
using the condition $b(r_0)=r_0$. At the throat, we
have $b'(r_0) = \frac{6 \pi }{\beta +9 \pi }-1 <1$. This imposes the restriction  $-6 \pi <\beta <-3 \pi$ for which $0< b'(r_0)<1$. Moreover, we verify that $\frac{b(r)}{r}\to 0$ as $r\to \infty$ which confirms the asymptotic flatness. Using the Eq. (\ref{bb3}), we plot the shape function for $\beta =-4.5 \pi$ (dot-dashed),  
$\beta = -5\pi$ (dashed) and  $\beta =-5.5 \pi$ (dotted) in the left panel of Fig. \ref{fig1}, when $r_0=1$. 

\begin{figure*}[h]
    \centering
    \includegraphics[width = 5.8 cm,height=5.8cm]{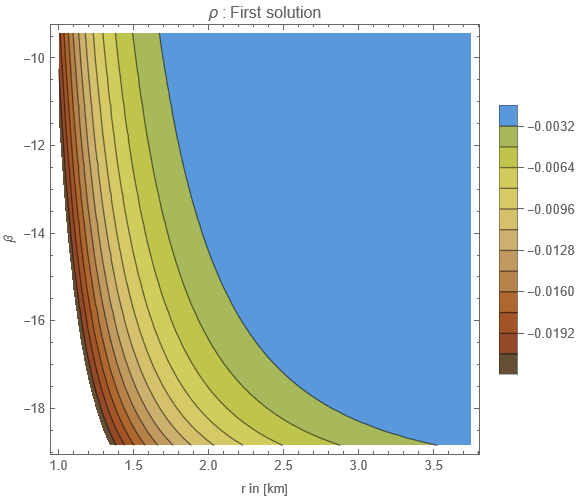} 
    \includegraphics[width = 5.8 cm,height=5.8cm]{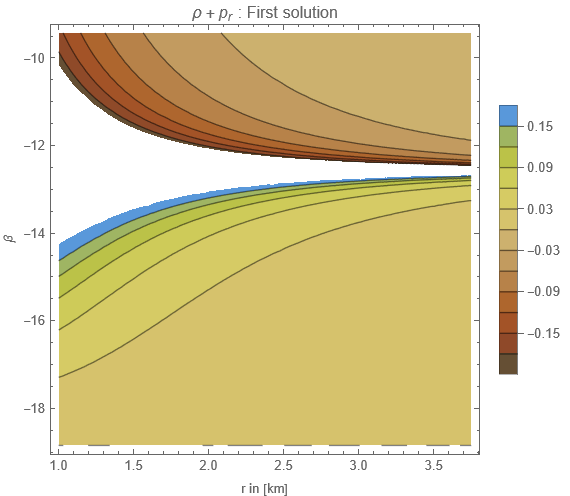} 
    \includegraphics[width = 5.8 cm,height=5.8cm]{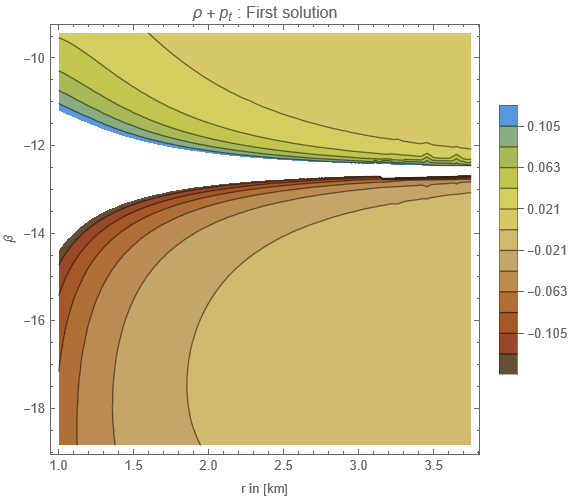} 
    \caption{Here, we consider a specific redshift function $\Phi(r) = \log \left(1+\frac{r_0}{r}\right)$ and plots of the specific stress-energy profile for $\rho$ (Left panel), $\rho+p_r$ (Middle panel) and $\rho+p_t$ (Right panel) versus $r$. In all cases, we consider $\omega=1$, which reduces to a traceless EMT, with $T = 0$. See the text for specific choices of the considered parameters. }  \label{fig2}
\end{figure*}

To get a better understanding about the shape function, we find the equation for the embedding surface
 \begin{eqnarray}\label{embed}
\frac{dz}{dr} = \pm \left(\frac{r}{b(r)}-1\right)^{-1/2}.
\end{eqnarray}

In Fig. \ref{fig1} (middle and right panels), we present the embedding diagram and embedded surface $z(r)$ with Eqs. (\ref{bb3}) and (\ref{embed}). The location of
the WH throat is defined by the black ring that connects two distinct universes.

Using Eqs. (\ref{eq27})-(\ref{eq28}) and  Eq. (\ref{bb3}) one finds the following relationships
\begin{eqnarray}
\rho &=& -\frac{3 r_0^2}{8 (\beta +9 \pi ) r^4}u(r) , \label{eq32}\\
\rho+ p_r &=& -\frac{r_0 \left(\beta  \left(r^2+r_0^2\right)+6 \pi  r^2\right) u(r)+2 (\beta +9 \pi ) r^2 r_0}{2 (\beta +4 \pi ) (\beta +9 \pi ) r^4 (r+r_0)}, \label{eq33}\\ 
\rho+p_t &=& \frac{r_0 \left((\beta+6 \pi)  \left(r^2-3 r r_0-2 r_0^2\right)  \right) u(r)+2 (\beta +9 \pi ) r^2 r_0}{4 (\beta +4 \pi ) (\beta +9 \pi ) r^4 (r+r_0)}, \label{eq34}
\end{eqnarray}
where $u(r)=\left(\frac{\beta  r+6 \pi  r+3 \pi  r_0}{\beta  r+9 \pi  r}\right)^{\frac{\beta }{3 \pi }}$. For $r> r_0=1$, we plot $\rho$, $\rho+p_r$ and $\rho+p_t$ as functions of $r$ in the Fig. \ref{fig2}.  It is clear that both $\rho+p_r$ and $\rho+p_t$ tend to zero as $r \to \infty$. Clearly, for $-6 \pi <\beta <-3 \pi$, energy density is negative outside the throat, while $\rho+p_r >0$ and $\rho+p_t <0$. From Fig. \ref{fig2} the NEC is violated outside the throat. The parameters of the solution are $-6 \pi <\beta <-3 \pi$ and $r_0=1$, respectively.

At the throat, $r=r_0$, one has
\begin{eqnarray} 
&&(\rho+ p_r)|_{r_0}= -\frac{\beta +6 \pi }{(\beta +4 \pi ) (\beta +9 \pi ) r_0^2},\\
&&(\rho+p_t)|_{r_0}= -\frac{\beta }{4 (\beta +4 \pi ) (\beta +9 \pi ) r_0^2}
\end{eqnarray}
It is obvious that $(\rho+ p_r)|_{r_0} \gtrless 0$ depending on the values of $-6 \pi <\beta < -3 \pi$. In the region $-4 \pi <\beta < -3 \pi$ we have $(\rho+ p_r)|_{r_0} < 0$,  and $(\rho+ p_r)|_{r_0} > 0$ for $-6 \pi <\beta < -4 \pi$. Moreover, we see that the situation is just reversed for the NEC along transverse direction (at the throat) i.e., $(\rho+ p_t)|_{r_0} \lessgtr 0$. Note that equations are not-valid for $\beta = -4 \pi$. From these observations, we can infer that the model violates the NEC throughout the spacetime and consequently the violation of WEC  also occurs.

\begin{figure}[h]
    \centering
    \includegraphics[width = 8.4 cm,height=5.8cm]{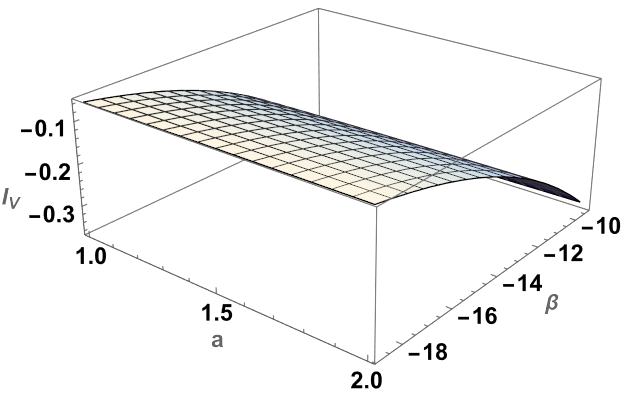}
    \caption{The plot depicts the total amount of exotic matter using the Eq. (\ref{eq38a}). This graph shows that for $I_V \to 0$, when taking the limit $a \to r_0$.}  \label{fig3}
\end{figure}

Here, we are also interested in quantifying the total amount of exotic matter through a \enquote{volume integral quantifier} (VIQ) used by Kar aand Sahdev\cite{Kar:2004hc}. This can be achieved through the definite integral (with a cut-off of the stress-energy at $a>r_0$): 
\begin{equation}\label{VIQ}
I_V = \int\left(\rho+p_r\right)dV= 2\int^{a}_{r_0}\left(\rho+p_r\right)4\pi r^2 dr,
\end{equation}
and using the Eq. (\ref{eq33}), we have
\begin{widetext}
\begin{eqnarray}\label{eq38a}
&& I_V =  \left. \frac{4 \pi  r_0 u(r) v(r) \left(3 \pi  n(r) r l(r)-6 \pi  (\beta +3 \pi )^2 r (v(r))^{-1} m(r)+\beta ^2 p(r) (v(r))^{-1}\right)}{\beta  (\beta +3 \pi ) (\beta +4 \pi ) (\beta +9 \pi ) r}  -\frac{8 \pi r_0  \log (r+r_0)}{(\beta +4 \pi ) }  \right|_{r_0}^{a},
\end{eqnarray}
\end{widetext}
 where,
\begin{eqnarray}
    v(r)&=&\left(\frac{(\beta +6 \pi ) r}{3 \pi  r_0}+1\right)^{-\frac{\beta }{3 \pi }},\\ l(r)&=&\, _2F_1\left(-\frac{\beta }{3 \pi },-\frac{\beta }{3 \pi };1-\frac{\beta }{3 \pi };-\frac{r (\beta +6 \pi )}{3 \pi  r_0}\right),\\ m(r)&=&\, _2F_1\left(-\frac{\beta }{3 \pi },-\frac{\beta }{3 \pi };1-\frac{\beta }{3 \pi };-\frac{r (\beta +6 \pi )}{3 \pi  r_0}\right),\\ n(r)&=&\beta ^2+9 \pi  \beta +18 \pi ^2,\\ p(r)&=&\beta  r+3 \pi  (2 r+r_0),
\end{eqnarray} 
After carefully reviewing the Eq. (\ref{eq38a}), we can conclude that $I_V = \int\left(\rho+p_r\right)dV \to 0$, when taking the limit $a \to r_0$.  For more clarity, we have plotted Fig. \ref{fig3}, which depicts that one can theoretically construct a traversable WH with small amount of NEC violating matter within the specified range of $-6 \pi <\beta <-3 \pi$.
\subsection{Solution for $b(r) = \frac{r_0^2}{r}$} 

\begin{figure*}[h]
    \centering
    \includegraphics[width = 5.8 cm,height=5.8cm]{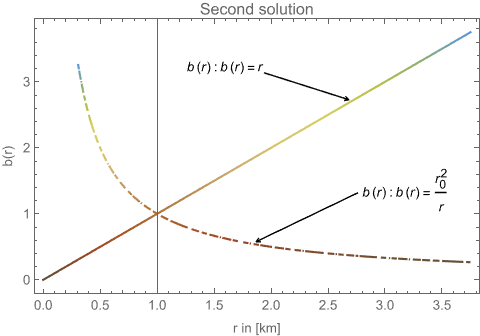} 
    \includegraphics[width = 5.8 cm,height=5.8cm]{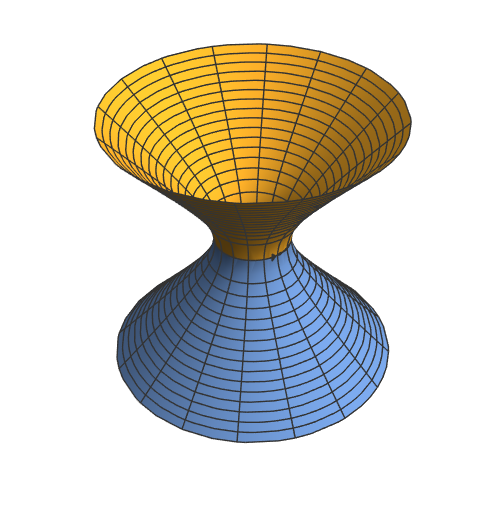} 
    \includegraphics[width = 5.8 cm,height=5.8cm]{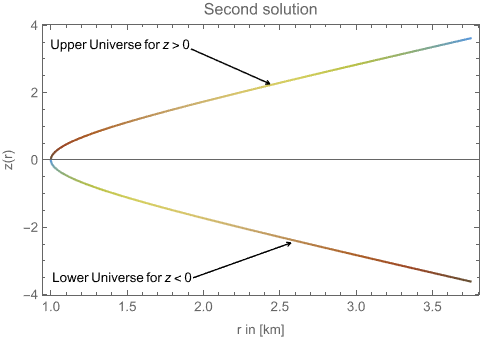} 
    \caption{Left panel: The behavior of $b(r) = r$ (solid black), and other curves are plotted for $b(r)$ given by $b(r) =\frac{r_0^2}{r}$ which shows $b(r) < r$. Middle panel:  Using the shape function $b(r) =\frac{r_0^2}{r}$ and Eq. (\ref{embed}), we plot the embedded diagram, wich provide valuable insights into the geometry and topology of the WH configurations. Right panel: Using the shape function $b(r) =\frac{r_0^2}{r}$ and Eq. (\ref{embed}), we plot the embedded surface $z(r)$, which provide valuable insights into the geometric structure and characteristics of the embedded surface. In all cases, we consider $\omega=1$, which reduces to a traceless EMT, with $T = 0$.  See the text for specific choices of the considered parameters. }  \label{fig4}
\end{figure*}

 Here, we consider a widely used specific form function  given by $b(r) = \frac{r_0^2}{r}$. Notably a generalised version of this shape function appears in the work of Tripathy \cite{tripathy}. However, in that work the author commenced the study by referring to  `tideless' forces amounting to $\phi = 0$. Thereafter he adopted the classical Casimir force variation density going as $-\frac{1}{a^4}$ where $a$ is the separation between two charged plates. Based on these he obtained the shape function above. In addition he was interested in quantum effects and the generalised uncertainty principle (GUP). Zubair {\it {et al}} \cite{zubair2} followed a similar approach to Tripathy however to fix the geometry of the models they assumed the existence of conformal Killing vectors. This places a restriction on the metric potentials, that is on the shape function profile.  Our approach is quite different: We take the conventional understanding of vanishing tidal forces as $\frac{d\phi}{dr}  = 0$ and then we postulate forms of the shape function with the desired properties for a viable WH. In the Fig. \ref{fig4}, we plot the shape function $b(r)$ along with the embedding diagram and the embedded surface $z(r)$ using $b(r) =\frac{r_0^2}{r}$ and Eq.  (\ref{embed}). 
 
 Taking into account Eqs. (\ref{12})-(\ref{14}) with the EoS (\ref{EoS1}), we have the following expressions
\begin{eqnarray}
 -\frac{r_0^2}{r^4} &=&\frac{8}{3} [\beta +3 \pi ] [p_r+2 p_t],\label{eq37}\\
\frac{2 \left(r^3-r r_0^2\right) \Phi '-r_0^2}{r^4}&=& \frac{4}{3} [(\beta +6 \pi ) p_r-\beta  p_t], \label{eq38} \\
 \frac{\left(r^4-r^2 r_0^2\right) \left(\Phi ''+(\Phi ')^2\right) +r^3 \Phi '+r_0^2}{r^4} 
  &=& \frac{2}{3} [(\beta +12 \pi ) p_t-\beta  p_r]. \label{eq39} 
\end{eqnarray}
Solving the Eqs. (\ref{eq37})-(\ref{eq38}), we get the  expression for $p_r$ and $p_t$ in terms  of $b$ and $b^{\prime}$, 
\begin{align}
 p_r(r)&=\frac{8 (\beta +3 \pi ) r (r-r_0) (r+r_0) \Phi '-(5 \beta +12 \pi ) r_0^2}{8 (\beta +3 \pi ) (\beta +4 \pi ) r^4},\label{eq40}\\
 p_t(r)&= \frac{4 (\beta +3 \pi ) r (r_0-r) (r+r_0) \Phi '+\beta  r_0^2}{8 (\beta +3 \pi ) (\beta +4 \pi ) r^4},\label{eq41}
\end{align}
Substituting $p_r$ and $p_t$ into Eq. (\ref{eq39}), we
obtain the following relationship
\begin{eqnarray} \label{eq42}
2 \Phi ' \left(2 r^2+r (r-r_0) (r+r_0) \Phi '-r_0^2\right) 
+\frac{(\beta +6 \pi ) r_0^2}{(\beta +3 \pi ) r} +2 r (r-r_0) (r+r_0) \Phi ''(r)=0,
\end{eqnarray}
which may be integrated to yield the solution
\begin{eqnarray} \label{eq43}
\Phi(r) = \ln \left(\cos \left(\frac{\mathcal{A}}{{\sqrt{2} r_0}} \cot ^{-1}\left(\frac{r_0}{\sqrt{r^2-r_0^2}}\right)-c_1\right)\right)+c_2 \nonumber \\
\end{eqnarray}
where $\mathcal{A} =\sqrt{\frac{(\beta +6 \pi ) r_0^2}{\beta +3 \pi }}$ and with $c_1$ and $c_2$ being integration constants, which at the throat, reduce to
\begin{eqnarray} \label{eq44}
\Phi(r_0) = \ln \left(\cos \left(c_1\right)\right)+c_2, 
\end{eqnarray}
We need to impose the condition $\cos\left(c_1\right) >0$ to ensure a regular solution. Thus,  $\Phi(r)$ is finite at the throat. Moreover, we know that the range of $\cos(x)$ is $ -1\leq \cos(x) \leq 1$, and $\ln(x)$ exists only when $x >0$. Then the proposed range of  $0< \cos (x)=\cos \left(\frac{\mathcal{A}}{{\sqrt{2} r_0}} \cot ^{-1}\left(\frac{r_0}{\sqrt{r^2-r_0^2}}\right)-c_1\right)\leq 1$ for which $\ln(x)$ is defined. If we consider the maximum value of $\cos \left(x\right)=1$ then $\ln(x) =0$.  Moreover, $\cot ^{-1}\left(\frac{r_0}{\sqrt{r^2-r_0^2}}\right) \to \cot ^{-1}(0) =\frac{\pi}{2}$ as $r\to \infty $. This gives $\ln\left( \cos \left(\frac{\pi}{2} \frac{\mathcal{A}}{{\sqrt{2} r_0}} -c_1\right)\right)$ a finite value, as the range of $ \cos (x) $ is finite. Finally, we arrive at the conclusion that $\Phi(r)$ given in Eq. (\ref{eq43}) is finite as $r\to \infty $ so that this solution now reflects a traversable WH.

\begin{figure*}[h]
    \centering
    \includegraphics[width = 5.8 cm,height=5.8cm]{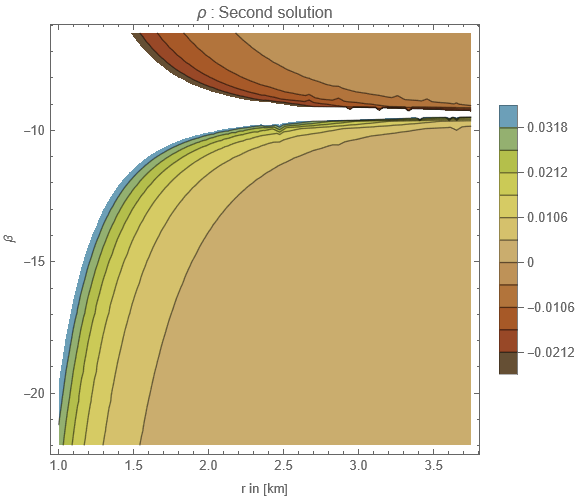} 
    \includegraphics[width = 5.8 cm,height=5.8cm]{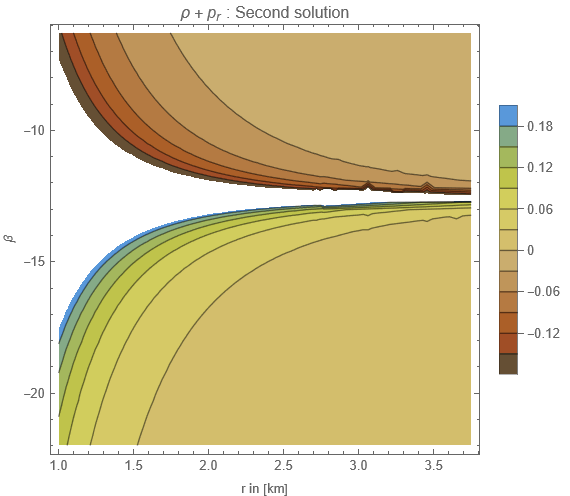} 
    \includegraphics[width = 5.8 cm,height=5.8cm]{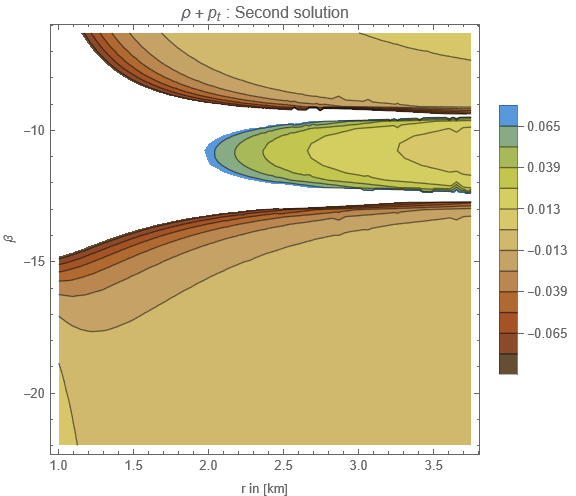} 
    \caption{: Here, we consider a specific shape function $b(r) =\frac{r_0^2}{r}$ and plots of the specific stress-energy profile for $\rho$ (Left panel), $\rho+p_r$ (Middle panel) and $\rho+p_t$ (Right panel) versus $r$. In all cases, we consider $\omega=1$, which reduces to a traceless EMT, with $T = 0$. See the text for specific choices of the considered parameters. }  \label{fig6}
\end{figure*}

Considering the redshift function (\ref{eq43}) and inserting the field equations (\ref{eq40})-(\ref{eq41})   into  Eq. (\ref{EoS1}), we obtain 
\begin{align}
\rho &= -\frac{3 r_0^2}{8 (\beta +3 \pi ) r^4}, \label{eq46}\\
\rho+ p_r &= \frac{\sqrt{2} \sqrt{(r^2-r_0^2)} \mathcal{A} \tan\theta -2 r_0^2}{2 (\beta +4 \pi ) r^4}, \label{eq47}\\ 
\rho+p_t &= -\frac{\sqrt{2} \sqrt{(r^2-r_0^2)} \mathcal{A} \tan \theta +\mathcal{A}^2}{4 (\beta +4 \pi ) r^4}, \label{eq48}
\end{align}
where $\mathcal{A}= \sqrt{\frac{(\beta +6 \pi ) r_0^2}{\beta +3 \pi }}$
and $\theta= \frac{1}{\sqrt{2} r_0}\left(\sqrt{2} r_0 c_1-\mathcal{A} \cot ^{-1}\left(\frac{r_0}{\sqrt{r^2-r_0^2}}\right)\right)$.  In Fig. \ref{fig5}, we plot $\rho$, $\rho+p_r$ and $\rho+p_t$, respectively.  For this case, we have plotted a diagram considering different values of $\beta$. From the Fig. \ref{fig4}, we see that the energy density, and two components for $\rho+p_r$ and $\rho+p_t$ are negative outside 
the throat i.e., $r> r_0=1$. This means that the NEC is always violated outside the throat radius for $\beta > -3 \pi $. For the configurations with $\beta < -6 \pi $, it is interesting that the WEC (and also NEC) is satisfied, as can be seen from the right panel of Fig. \ref{fig4}. Note that $-6 \pi<\beta < -3 \pi $ is not an allowable range for WH construction from the form of $\mathcal{A}$.

The NEC at the throat is given by
\begin{align}
(\rho+ p_r)|_{r_0}  &= -\frac{1}{(\beta +4 \pi ) r_0^2}, \label{eq49}\\
(\rho+ p_t)|_{r_0} &= -\frac{\beta +6 \pi }{4 (\beta +3 \pi ) (\beta +4 \pi ) r_0^2}. \label{eq50}
\end{align}
It is clear that both   $(\rho+ p_{r,t})|_{r_0} <0$ when $\beta > -3 \pi $. Thus, the NEC is always
violated for $\beta > -3 \pi $. On the other hand, we can verify  that $(\rho+ p_{r,t})|_{r_0} >0$ for $\beta < -6 \pi $.
This would imply that $\rho$, $\rho+p_r$  and  $\rho+p_t$ are always positive when $\beta < -6 \pi $. As a closing remark we can say that realistic physical mechanisms such as the Casimir effect exist in order to support the geometry of traversable WHs. Exotic matter fields in the form of dark energy are therefore not compulsory. This is important as experimental support for dark energy is absent to date, whereas the Casimir phenomenon is a laboratory proven counter-intuitive effect.   

We now consider the VIQ to evaluate the total amount of exotic matter. With the help of Eqs. (\ref{eq47}) and (\ref{VIQ}), we are in a position to discuss the amount of exotic matter. The full analytical integration is not feasible on account of the complicated nature of the expressions. Thus, we  perform 
numerical integration for specific values of $\beta = -2 \pi $ and $c_1=1$ with $\tan \theta \approx \theta$ for small values of $\theta$. Finally, we have
\begin{figure}[h]
    \centering
    \includegraphics[width = 8.5 cm,height=5.8cm]{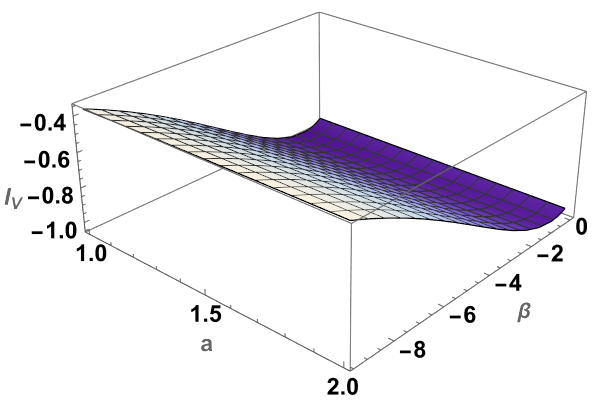}
    \caption{The plot depicts the total amount of exotic matter using the Eq. (\ref{eq47}). This graph shows that for $I_V \to 0$, when taking the limit $a \to r_0$.}  \label{fig6}
\end{figure}

\begin{equation}\label{eq52a}
	I_V =  2\int^{a}_{r_0}\left(\rho+p_r\right)4\pi r^2 dr \\
\approx \frac{\sqrt{2} \left(a^2-r_0^2\right)^{3/2} \mathcal{A}}{6 (\beta +4 \pi ) a^3 r_0^2}.
\end{equation}
As a result, we clearly see that $I_V  \to 0$ when $a\to r_0$. For greater transparency, we plot Fig. \ref{fig6} for the range of $-3\pi<\beta<0$ where we find that $\rho+p_r<0$.


\subsection{Solution for $b(r) = r_0+\gamma  r_0 \left(1-\frac{r_0}{r}\right)$} 

\begin{figure*}[h]
    \centering
    \includegraphics[width = 5.8 cm,height=5.8cm]{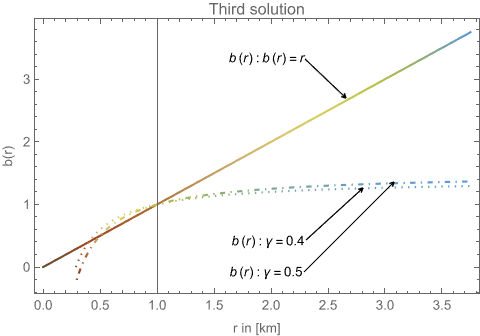} 
    \includegraphics[width = 5.8 cm,height=5.8cm]{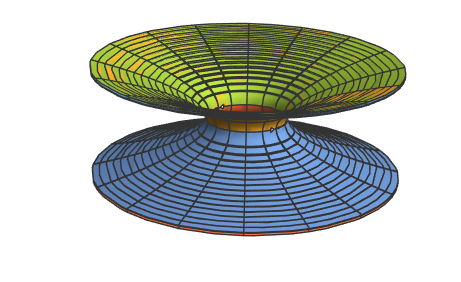} 
    \includegraphics[width = 5.8 cm,height=5.8cm]{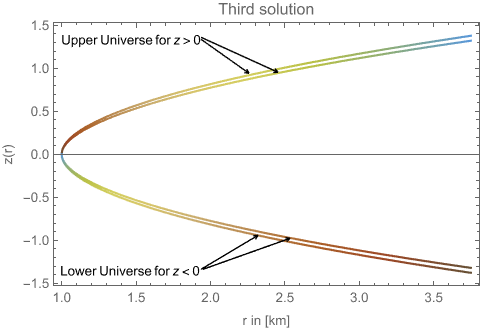} 
    \caption{Left panel: The behavior of $b(r) = r$ (solid black), and other curves are plotted for $b(r)$ given by $b(r) =\frac{r_0^2}{r}$ which shows $b(r) < r$. Middle panel:  Using the shape function $b(r) = r_0+\gamma  r_0 \left(1-\frac{r_0}{r}\right)$ and Eq. (\ref{embed}), we plot the embedded diagram, wich provide valuable insights into the geometry and topology of the WH configurations. Right panel: Using the shape function $b(r) = r_0+\gamma  r_0 \left(1-\frac{r_0}{r}\right)$ and Eq. (\ref{embed}), we plot the embedded surface $z(r)$, which provide valuable insights into the geometric structure and characteristics of the embedded surface. In all cases, we consider $\omega=1$, which reduces to a traceless EMT, with $T = 0$.  See the text for specific choices of the considered parameters. }  \label{fig7}
\end{figure*}

Here, we consider the specific model for $b(r) = r_0+\gamma  r_0 \left(1-\frac{r_0}{r}\right)$ \cite{Lobo:2008zu}, 
where $0 < \gamma< 1$  is of particular interest in generating WH solutions satisfying the condition $b'(r_0)<1$. In Fig. \ref{fig7}, we have plotted the shape function $b(r)$ along with the embedding diagram and embedded surface $z(r)$ corresponding to the above-mentioned shape function. Now, we take the field equations (\ref{12})-(\ref{14}) with the EoS (\ref{EoS1}), which gives
\begin{widetext}
\begin{eqnarray}
\frac{\gamma  r_0^2}{r^4}&=&\frac{8}{3} (\beta +3 \pi ) (p_r+2 p_t),\label{eq50}\\
 k(r) r \Phi '+ l(r)&=& -4 (\beta +6 \pi ) r^4 p_r+4 \beta  r^4 p_t,    \label{eq51} \\
 k(r) r \Phi ''+6 r^2 (r-r_0) (r-\gamma  r_0) \Phi '^2+3 r^2 (2 r-(\gamma +1) r_0) \Phi '+l(r) &=& -4 \beta  r^4 p_r+ 4 (\beta +12 \pi ) r^4 p_t. \nonumber \\\label{eq52}
\end{eqnarray}
\end{widetext}
where $k(r)= 6 r (r-r_0) (r-\gamma  r_0)$ and $l(r)=3 r_0 (\gamma  r+r-\gamma  r_0)$. Equations (\ref{eq50}) and (\ref{eq51}) enable us to obtain  explicit  expressions for $p_r$ and $p_t$ in terms  of $b$ and $b^{\prime}$,
\begin{widetext}
\begin{eqnarray}
&& p_r= \frac{8 (\beta +3 \pi ) r (r-r_0) (r-\gamma  r_0) \Phi '+\beta  r_0 (5 \gamma  r_0-4 (\gamma +1) r)-12 \pi  r_0 (\gamma  r+r-\gamma  r_0)}{8 (\beta +3 \pi ) (\beta +4 \pi ) r^4}, \label{eq53}\\
&& p_t= \frac{r_0 (2 \beta  (\gamma +1) r+6 \pi  (\gamma +1) r-\beta  \gamma  r_0)-4 (\beta +3 \pi ) r (r-r_0) (r-\gamma  r_0) \Phi '}{8 (\beta +3 \pi ) (\beta +4 \pi ) r^4}. \label{eq54}
\end{eqnarray}
\end{widetext}
Inserting $p_r$ and $p_t$ into  Eq. (\ref{eq52}), one
obtains the following relationship
\begin{align}
 \Phi ' \left(4 r^2+2 r (r-r_0) (r-\gamma  R) \Phi '-3 (\gamma +1) r r_0+2 \gamma  r_0^2\right) \nonumber\\
 +2 r (r-r_0) (r-\gamma  r_0) \Phi '' =\frac{(\beta +6 \pi ) \gamma  r_0^2}{(\beta +3 \pi ) r},
\end{align}
which may be integrated to yield the solution
\begin{align}\label{eq56}
\Phi (r)= \log \left(\cos \left(\frac{\sqrt{2} \mathcal{K} \tanh ^{-1}\left(\frac{\sqrt{\gamma } \sqrt{r-r_0}}{\sqrt{r-\gamma  r_0}}\right)}{\sqrt{\gamma } r_0}-k_1\right)\right)+k_2
\end{align}
where $\mathcal{K}=\sqrt{-\frac{(\beta +6 \pi ) \gamma  r_0^2}{\beta +3 \pi }}$ with $k_1$ and $k_2$ are integrating constants, which at the throat, reduce to
\begin{eqnarray} \label{eq44}
\Phi(r_0) = \ln \left(\cos \left(k_1\right)\right)+k_2.
\end{eqnarray}
At this juncture we point  out a few considerations to maintain the regularity of the solution.  We need to impose $\cos\left(k_1\right) >0$, so that $\Phi(r)$ is finite at the throat. Interestingly, we see that the present case is  similar to our previous case, and thus we do not repeat our discussion here. We may conclude that
$\Phi(r)$ given in Eq. (\ref{eq56}) is finite as $r\to \infty $, which permits  a horizonless solution. 

\begin{figure*}[h]
    \centering
    \includegraphics[width = 5.8 cm,height=5.8cm]{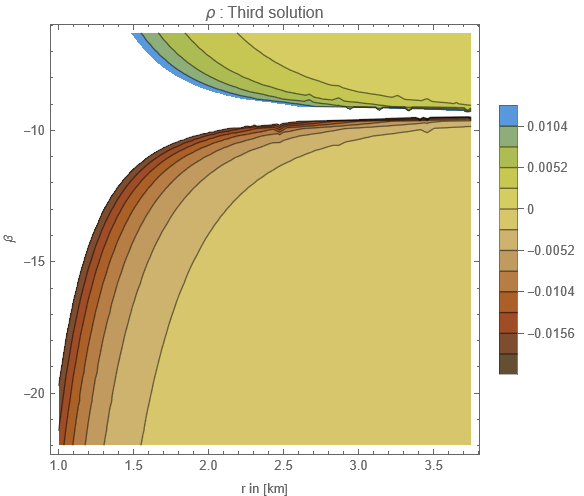} 
    \includegraphics[width = 5.8 cm,height=5.8cm]{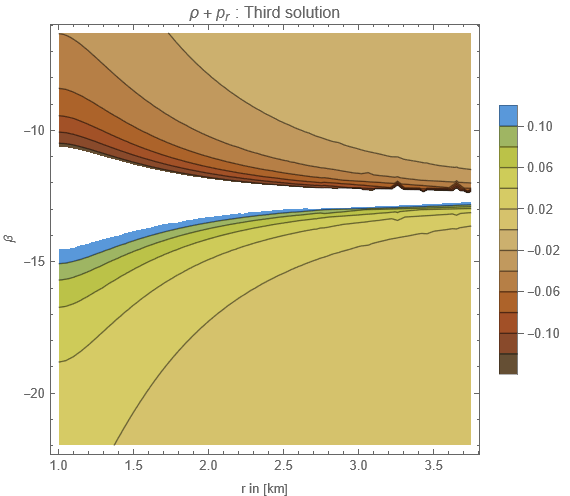} 
    \includegraphics[width = 5.8 cm,height=5.8cm]{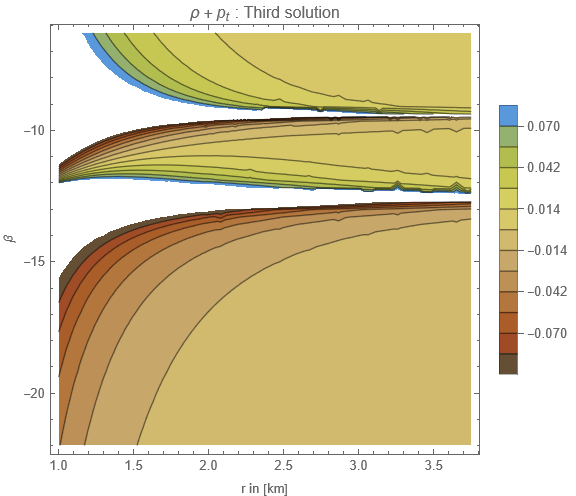} 
    \caption{: Here, we consider a specific shape function $b(r) = r_0+\gamma  r_0 \left(1-\frac{r_0}{r}\right)$ and plots of the specific stress-energy profile for $\rho$ (Left panel), $\rho+p_r$ (Middle panel) and $\rho+p_t$ (Right panel) versus $r$. In all cases, we consider $\omega=1$, which reduces to a traceless EMT, with $T = 0$. See the text for specific choices of the considered parameters. }  \label{fig8}
\end{figure*}

Inserting the redshift function (\ref{eq56}) into the expressions (\ref{eq53})-(\ref{eq54}) and using  Eq. (\ref{EoS1}), we finally have
\begin{widetext}
\begin{align}
\rho &= \frac{3 \gamma  r_0^2}{8 (\beta +3 \pi ) r^4}, \label{eq58}\\
\rho+ p_r &= \frac{4 \sqrt{2} (\beta +3 \pi )  r \sqrt{r-r_0} \mathcal{M}(r) +\beta  r_0 (5 \gamma  r_0-4 (\gamma +1) r)-12 \pi  r_0 (\gamma  r+r-\gamma  r_0)}{8 (\beta +3 \pi ) (\beta +4 \pi ) r^4}, \label{eq59}\\ 
\rho+p_t &= \frac{r_0 (2 \beta  (\gamma +1) r+6 \pi  (\gamma +1) r-\beta  \gamma  r_0-2 \sqrt{2} (\beta +3 \pi ) \mathcal{M}(r) \sqrt{r-r_0}}{8 (\beta +3 \pi ) (\beta +4 \pi ) r^4}, \label{eq60}
\end{align}
\end{widetext}
where $\mathcal{M}(r)= \sqrt{r-\gamma  r_0} \mathcal{K} \tan \left(k_1-\frac{\sqrt{2} \mathcal{K} \tanh ^{-1}\left(\frac{\sqrt{\gamma } \sqrt{r-r_0}}{\sqrt{r-\gamma  r_0}}\right)}{\sqrt{\gamma } r_0}\right)$. Having obtained the solutions for the matter fields, we plot $\rho$, $\rho+p_r$ and $\rho+p_t$ in Fig. \ref{fig8}. It is important to emphasize that the region for the domain is $-6 \pi<\beta < -3 \pi $ where the 
WH is valid; see the expression $\mathcal{K}$. Note also that $\beta= -4\pi $  and $\beta= -3 \pi $ is a point of discontinuity where the solution is not valid. Fig. \ref{fig8} depicts the behaviour of the stress-energy components
for $-6 \pi<\beta < -3 \pi $ and $\gamma=0.5$. Observing the figure, we see that $\rho$ and $\rho+p_r$ are negative, while  $\rho+p_t$ is positive outside the throat i.e., $r> r_0=1$. This clearly indicate that the NEC is violated  outside  the throat radius  within the specified region. 

The NEC at the throat is given by
\begin{align}
(\rho+ p_r)|_{r_0}  &= \frac{\gamma -1}{2 (\beta +4 \pi ) r_0^2}, \label{eq61}\\
(\rho+ p_t)|_{r_0} &= \frac{2 \beta  \gamma +\beta +\pi  (9 \gamma +3)}{4 (\beta +3 \pi ) (\beta +4 \pi ) r_0^2}. \label{eq62}
\end{align}
In this case, we see that $(\rho+ p_{r})|_{r_0} <0$ and  $(\rho+ p_{t})|_{r_0} \lessgtr 0$ when $-4 \pi<\beta < -3 \pi $.
On the other hand,  $(\rho+ p_{r})|_{r_0} >0$ and  $(\rho+ p_{t})|_{r_0} < 0$ when $-6 \pi<\beta < -4 \pi $.
Thus, it is clear that the NEC is always violated. 
\begin{figure}[h]
    \centering
    \includegraphics[width = 8.4 cm,height=5.8cm]{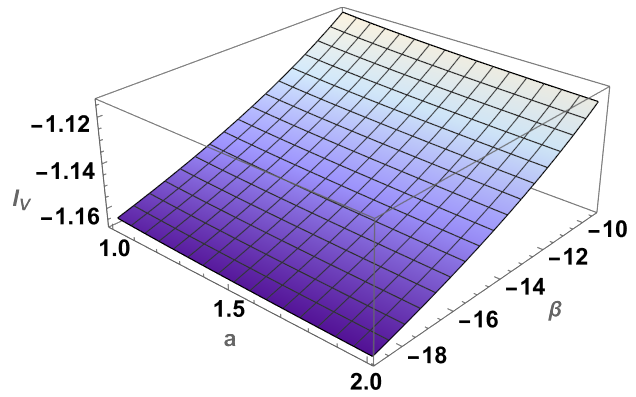}
    \caption{The plot depicts the total amount of exotic matter using the Eq. (\ref{eq59}). This figure clearly indicate that the total amount of exotic matter can be negligible depending on the model parameters.}  \label{fig9}
\end{figure}

Finally, we determine the total amount of exotic matter through the VIQ. Thus, using  Eq. (\ref{VIQ}) and Eq. (\ref{eq66}), 
we obtain an approximate analytical solution for the definite integral with a cut-off of the stress-energy at $a>r_0$: 
\begin{eqnarray}\label{eq66}
    I_V &=& \frac{1}{{48 (\beta +4 \pi ) \gamma ^{5/2} a^3 r_0^2}} \left[ \sqrt{\gamma } \left( \left(3 \gamma ^2-2 \gamma +3\right) a^2 l_1 \right. \right.  \left.\left. -8 \gamma ^2 r_0^2 \left(l_1+2 \gamma  r_0^2\right)  +2 \gamma  (\gamma +1) a r_0 \left(l_1+6 \gamma  r_0^2\right)\right)  \right. \nonumber\\
    && \left.  -3 \sqrt{2} (\gamma -1)^2 (\gamma +1) a^3 l_2 \right],
\end{eqnarray}
where $l_1=\sqrt{2}\sqrt{a-r_0} \sqrt{a-\gamma  r_0}$ and $l_2=\tanh ^{-1}\left(\frac{a-\sqrt{r-r_0} \sqrt{a-\gamma  r_0}}{\sqrt{\gamma } r_0}\right)$.  Fig. \ref{fig9} is illustrated  for $\gamma=0.5$, $k_1=1$ and varying $\beta$ in the domain  $-6 \pi<\beta < -3 \pi $  where the WH solution is valid. Observing Fig. \ref{fig9} we may infer that  the amount of exotic matter can be
made arbitrarily small through a judicious choice of the parameter space.  

\begin{figure*}[h]
    \centering
    \includegraphics[width = 5.8 cm,height=5.8cm]{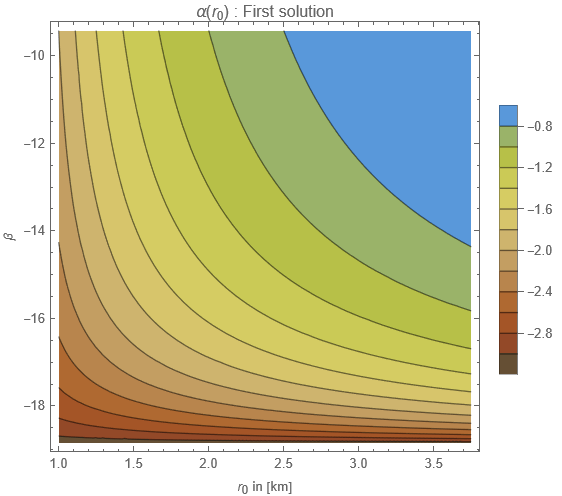} 
    \includegraphics[width = 5.8 cm,height=5.8cm]{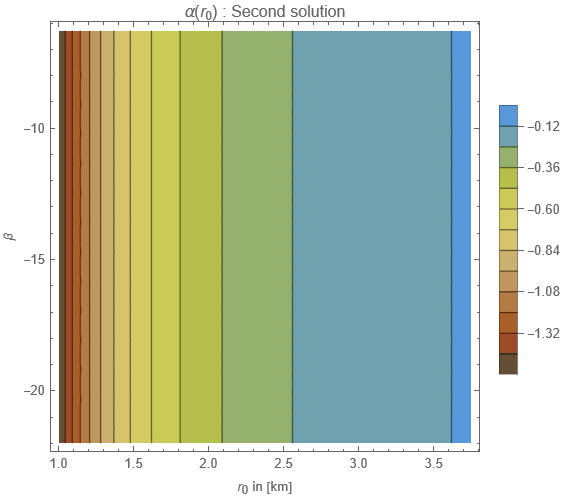} 
    \includegraphics[width = 5.8 cm,height=5.8cm]{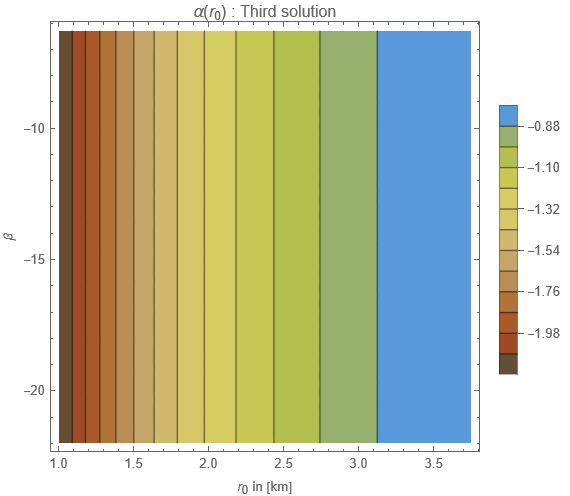} 
    \caption{ The plots depict the behavior of the numerical results of the deflection angle of photons, represented by the quantity $\alpha(r_0)$, for three solutions. These plots demonstrate the negative deflection angle, indicating the repulsive nature of gravity in three solutions. In all cases, we consider $\omega=1$, which reduces to a traceless EMT, with $T = 0$.  See the text for specific choices of the considered parameters. }  \label{fig10}
\end{figure*}

\section{The deflection angle of photons along null geodesics}\label{sec6}

To explore the deviation angle of photons from null geodesics, let us begin by introducing a comprehensive line element that embodies both spherically symmetric and static characteristics \cite{misner1973, schutz2014}. This line element can be represented as follows,
\begin{equation}\label{eq38}
	ds^2 = -H(r) dt^2 +G(r) dr^2 +F(r) d\Omega^2.
\end{equation}

To characterize the trajectory of an object in free fall with respect to the underlying geometry \cite{schutz2014}, we make use of the geodesic equation. This equation establishes a relationship between the momentum one-forms of the object and the spacetime's geometry. It can be mathematically expressed as follows,
\begin{equation}\label{eq39}
	\frac{dp_\beta}{d\lambda} = \frac{1}{2} g_{\nu \alpha, \beta} p^\nu p^\alpha .
\end{equation}
 In this scenario, we assign the parameter $\lambda$ to represent the affine parameter. It is important to emphasize that if the components of $g_{\alpha\nu}$ do not vary with respect to $x^\beta$ for a fixed index $\beta$, then $p_\beta$ remains constant throughout the motion. By considering exclusively the equatorial slice with $\theta = \pi/2$, we discover that all the components $g_{\alpha\beta}$ in Equation \eqref{eq39} become independent of $t$, $\theta$, and $\phi$. As a result, we can identify the corresponding Killing vector fields $\delta^{\mu}\alpha \partial\nu$ with $\alpha$ as a cyclic coordinate. By designating the constants of motion as $p_t$ and $p_\phi$, we can delve into the analysis of the system's dynamics using the following expressions,
\begin{equation}\label{eq40}
	p_t = -E, ~~~~~~~~~~~~~ p_\phi = L.
\end{equation}
Where $E$ and $L$ denote the energy and angular momentum of the photon, respectively. With this in mind, we can express the geodesic equation as,
\begin{eqnarray}\label{eq41}
p_t = \dot{t} = g^{t \nu} p_\nu = \frac{E}{H(r)}, \\
	p_\phi = \dot{\phi} = g^{\phi \nu} p_\nu = \frac{L}{F(r)},
\end{eqnarray}
Here, the overdot notation represents differentiation with respect to the affine parameter $\lambda$. Continuing from the previous analysis, we can straightforwardly obtain the radial null geodesic as,
\begin{equation}\label{eq42}
	\dot{r}^2 = \frac{1}{G(r)} \left[ \frac{E^2}{H(r)} - \frac{L^2}{F(r)} \right].
\end{equation}
To provide a more precise formulation, we can express the equation for the photon trajectory in terms of the impact parameter $\mu = L/E$ as,
\begin{equation}\label{eq43}
	\left[ \frac{dr}{d\phi} \right]^2 = \frac{F(r)^2}{\mu^2 G(r)} \left[ \frac{1}{H(r)} - \frac{\mu^2}{F(r)} \right].
\end{equation}
Now, let's consider a photon source with a radius $r_s$ that affects the underlying geometry. In order to determine the deflection angle of the photons, we need to identify the conditions under which the photons can reach the surface. This occurs when a solution $r_0$ satisfies the conditions $r_0 > r_s$ and $\dot{r}^2 = 0$, where $r_0$ represents the distance of closest approach or turning point. In this scenario, the impact parameter can be expressed as,
\begin{equation}\label{eq44}
	\mu = \frac{L}{E} = \pm \sqrt{\frac{F(r_0)}{H(r_0)}},
\end{equation}
In the regime of weak gravity, it becomes apparent that $ \mu \approx \sqrt{\mathcal{F}(r_0)} $. Therefore, if a photon originates from the polar coordinate limit defined as $ \lim\limits_{r \rightarrow \infty} \left( r, -\frac{\pi}{2}-\frac{\alpha}{2} \right) $, passes through the turning point located at $ (r_0, 0) $, and approaches $ \lim\limits_{r \rightarrow \infty} \left( r, \frac{\pi}{2}+\frac{\alpha}{2} \right) $, we can define the deflection angle of the photon as $ \alpha $. This deflection angle, denoted by $ \alpha $, is dependent on $r_0$ \cite{Bhattacharya:2010zzb}, and can be explicitly derived from Eq. (\ref{eq44}) as,
\begin{equation}\label{eq45}
	\alpha(r_0) = -\pi + 2 \int_{r_0}^{\infty}
	\sqrt{\frac{\mathcal{G}(r) }{\mathcal{F}(r)}}\left[ \left( \frac{\mathcal{H}(r_0)}{\mathcal{H}(r)} \right) \left( \frac{\mathcal{F}(r)}{\mathcal{F}(r_0)} \right) -1 \right]^{-1/2}dr.
\end{equation}
For the chosen metric coefficients in the $\mathcal{WH}$ geometry, one can now easily determine the deflection angle of photons in $f(R, T)$ gravity by numerically integrating the formulas mentioned above, taking into account the shape functions specified by Eqs. $b(r)= r_0 \left(\frac{  6\pi r+3 \pi r_0+ \beta r}{(\beta +9\pi)r }\right)^{\frac{3\pi+\beta}{3\pi}}$, $b(r) =\frac{r_0^2}{r}$, and $b(r) = r_0+\gamma  r_0 \left(1-\frac{r_0}{r}\right)$. Fig. \ref{fig10} displays the plots illustrating the phenomenon at hand. When the deflection angle is negative, it indicates the presence of repulsive gravity. To validate this phenomenon, we introduce the concept of the photon deflection angle on the WH. As stated in \cite{Panpanich:2019mll}, in the presence of repulsive gravity acting on photons, the deflection angle becomes negative in the spacetime. Remarkably, it is interesting to observe that the deflection angle consistently maintains negative values for all values of $r_0$ in the three solutions. This consistent negativity can be interpreted as a manifestation of the repulsive gravity effect.

\section{Concluding remarks}\label{sec7}

Invoking a Casimir type  together with suitable shape and redshift  functions, we presented  complete analytical solutions for static and spherically symmetric asymptotically flat WH geometries in the background of $f(R,T)$ theory.  The novelty of our work lies in the use of this Casimir stress energy which is known to produce negative density and which is a necessity for the emergence of WHs. The negative density causes a repulsive gravitational behaviour which sustains the opening of the WH throat. In contrast most treatments of WHs rely on the negative density arising out of exotic fluids such as dark energy. The Casimir effect has the advantage that it is laboratory confirmed whereas dark energy is a speculative entity lacking empirical observational evidence to date. Invoking the Casimir behaviour in the study of WH geometry has attracted only limited attention historically. In our work we have adopted the approach of Kar and Sahadev \cite{Kar:1995vm} which, to the best of our knowledge has not been attempted before. A similar endeavour by Tripathi \cite{tripathy} utilised the quantum effects variation of density with plate-distance as well as considering the generalised uncertainty principle.   

We have elected to study $f(R,T)$ gravity
which is a generalization of $f(R)$ gravity such that the gravitational Lagrangian density  depends 
on $R$ (Ricci scalar) and $T$ (stress-energy tensor). Part of the motivation of $f(R,T)$ gravity is the coupling between geometry and matter, which has extensive applications  in astrophysics and cosmology and the fact that this peculiar  has not featured in the literature in $f(R, T)$ theory. 

In the geometrical representation, we have analyzed different energy conditions and investigated  the effects of the coupling constant $\beta$ on the WH structure. At the outset we show that a sustainable WH could not be possible for a $\Phi^{\prime}(r) =0$ thereby ruling out WHs with vanishing tidal forces as is the case in the standard theory.  Subsequently  
by considering $\Phi(r) = \log \left(1+\frac{r_0}{r}\right)$, we show that viable WH solutions exist provided that the coupling parameter $\beta$ is constrained as $-6 \pi <\beta <-3 \pi$. In Fig. \ref{fig1}, we showed that the shape function $b(r)$ in Eq. (\ref{bb3}) satisfies all the necessary requirements, but the NEC is violated throughout the spacetime and consequently the violation of WEC also occurs. This is welcome in WH geometry and physics. Finally,  we consider $b(r) = \frac{r_0^2}{r}$ as a form function and derive the stress energy tensor components. In this process the obtained redshift function  $\Phi(r)$ in (\ref{eq43}) is finite everywhere. This is a clear indication that  spacetime has no event horizons. Moreover, we have shown that the WEC is satisfied throughout the entire spacetime by considering a negative value of coupling constant $\beta < -6 \pi $. In addition to our investigations into energy conditions, we also delved into the repulsive effect of gravity. Our findings revealed that the existence of repulsive gravity results in a negative deflection angle. To validate this phenomenon, we introduced the concept of the photon deflection angle on the WH. As documented in \cite{Panpanich:2019mll}, when photons are subjected to repulsive gravity, this deflection angle becomes negative within the spacetime. Importantly, it is worth noting that the deflection angle consistently maintains negative values across all values of $r_0$ in the three solutions. This consistent negativity can be interpreted as a manifestation of the repulsive gravity effect.

This study shows that the Casimir stress energy is an ideal candidate for the study of WHs and in future work we shall consider its effect in other modified gravitational field theories.

\begin{acknowledgments}
  In accordance with the visiting associateship scheme, A. Pradhan is grateful to IUCAA, Pune, India, for providing support and facilities. SH thanks the National Research Foundation of South Africa for support through Grant 138012. AE thanks the National Research Foundation of South Africa for the award of a postdoctoral fellowship.
\end{acknowledgments}\

\end{document}